\documentclass{article}

\usepackage[utf8]{inputenc}

\usepackage{biblatex}
\usepackage{graphicx}
\graphicspath{{./images/}}

\usepackage{hyperref}

\title{How Do AI Timelines Affect Existential Risk?}
\author{Stephen McAleese}
\date{August 2022}

\begin{document}

\maketitle

\begin{abstract}

Superhuman artificial general intelligence could be created this century and would likely be a significant source of existential risk. Delaying the creation of superintelligent AI (ASI) could decrease total existential risk by increasing the amount of time humanity has to work on the AI alignment problem. However, since ASI could reduce most risks, delaying the creation of ASI could also increase other existential risks, especially from advanced future technologies such as synthetic biology and molecular nanotechnology. If AI existential risk is high relative to the sum of other existential risk, delaying the creation of ASI will tend to decrease total existential risk and vice-versa. Other factors such as war and a hardware overhang could increase AI risk and cognitive enhancement could decrease AI risk. To reduce total existential risk, humanity should take robustly positive actions such as working on existential risk analysis, AI governance and safety, and reducing all sources of existential risk by promoting differential technological development.

\end{abstract}

\tableofcontents

\section{Introduction}

Recent progress in AI suggests that artificial general intelligence (AGI) that is as capable as humans on a wide variety of tasks is likely to be created this century. Once AGI exists, further improvement of its abilities would enable it to surpass human intelligence resulting in the creation of a superintelligent artificial general intelligence that is vastly more capable than humans at many important cognitive tasks.

A similar concept is “transformative AI” which is defined as “AI that precipitates a transition comparable to (or more significant than) the agricultural or industrial revolution” \cite{tai}. For the purpose of this report I’m going to use the acronym “ASI” (artificial superintelligence) as a short-hand for superintelligent general AI that is “much smarter than the best human brains in practically every field, including scientific creativity, general wisdom and social skills” \cite{bostrom1998}.

After an ASI is created, it would probably have a significant and long-lasting effect on human civilization and its trajectory. An ASI could also be a significant source of existential risk and could result in a highly undesirable outcome such as human extinction.

Although recently created AI systems such as GPT-3 can accomplish a wide variety of tasks, their understanding of the world, generality and performance is not high enough for them to be a significant existential risk to humanity.

To be a major source of existential risk, an AI system might need to have the kind of deep, cross-domain understanding humans have that enables them to significantly change the world by taking actions such as creating and implementing complex long-term plans or inventing powerful new technologies.

For example, an AI that wanted to invent and deploy advanced nanotechnology might need to have the ability to read and understand scientific papers, plan, carry out and interpret the results of experiments, and model the behavior of other actors such as humans.

Since current AI systems are not intelligent enough to have one or more of these general abilities or significantly transform the world, they are not a major source of existential risk. But as AI progress continues, humanity might someday create ASI systems that are intelligent and powerful enough to be an existential risk to humanity.

\subsection{Motivation}

Some philosophers believe that reducing existential risk could have extremely high expected value \cite{bostrom2013existential, stronglongtermism}. For example, axiological strong longtermism states that “impact on the far future is the most important feature of our actions today” and that “every option that is near-best overall is near-best for the far future” \cite{stronglongtermism}. Research on existential risk is also generally neglected \cite{bostrom2013existential}.

Since the creation of ASI may be a major source of existential risk, actions that reduce the level of existential risk posed by ASI would have extremely high expected value.

Researchers such as Ajeya Contra have analyzed when humanity might create transformative AI (TAI) in the future by, for example, using biological anchors to estimate the amount of computational power necessary for TAI and extrapolating past progress in AI to estimate when TAI will be created \cite{forecastingtai}. This kind of research is important because it informs humanity on how it should act to minimize AI risk. For example, AI safety research involving current machine learning methods is more likely to be relevant and valuable for TAI safety if shorter rather than longer timelines are expected. AI timelines could also affect funding decisions or other priorities.

Although a significant amount of research effort has been put into estimating AI timelines, apparently much less research has been directed at the question of how AI timelines affect existential risk and similarly which ASI arrival date would be most desirable from the standpoint of existential risk reduction. Most previous research seems to have been descriptive rather than normative with a focus on predicting the arrival date of ASI as if it were a fixed and inevitable moment in the future with less attention being directed at how AI timelines affect existential risk and which ASI arrival date would be most desirable given the goal of minimizing existential risk.

This report is focused on answering the latter question: how do AI timelines affect total existential risk? And given the goal of minimizing total existential risk, should we prefer a world where ASI is created in the near or far future?

Total existential risk can be defined as the cumulative probability of an existential catastrophe occurring over time. Since the creation of ASI is likely one of the main sources of existential risk facing humanity this century \cite{theprecipice}, it would be valuable to study how various AI development trajectories affect total existential risk.

\section{What is the magnitude of existential risk from ASI this century?}

Before we compare existential risks, it will be useful to estimate the existential risk contribution of ASI to total existential risk.

The amount of existential risk from ASI this century depends on how likely ASI is to be created this century and how likely an existential catastrophe is to occur afterwards. I’ll describe several sources of information we can use to estimate these two variables.

\subsection{Expert surveys}

A 2014 survey asked experts when they thought high-level machine intelligence (HLMI) would be created where HLMI was defined as a machine that can “carry out most human professions at least as well as a typical human.” The survey found a median estimate that HLMI had a 50\% chance of being developed by 2040. Respondents were also asked when they thought superintelligence would be developed which was defined as “machine intelligence that greatly surpasses the performance of every human in most professions.” The median estimate was a 10\% probability of superintelligence within 2 years after the creation of HLMI and a 75\% probability within 30 years. The same survey found that, on average, experts believed that HLMI had an 18\% chance of causing an existential catastrophe \cite{survey}.

Another survey \cite{aiimpactssurvey} estimated a 50\% probability of HLMI 50 years after 2016 and that the median expert believed the probability of an extremely bad outcome was 5\%.

\subsection{The Precipice}

In \textit{The Precipice} \cite{theprecipice}, Toby Ord estimates that the probability of unaligned AI causing an existential catastrophe is about 10\% in the 21st century.

\subsection{Metaculus}

The Metaculus prediction market currently predicts that there is a 50\% probability of artificial general intelligence being created by 2041 \cite{metaculus}.

\subsection{Ajeya Contra}

Ajeya Contra is a researcher at Open Philanthropy who has spent a significant amount of time predicting when transformative AI (TAI) will be created. Recently, she revised her median prediction from 2050 to 2040 \cite{twoyearupdate}.

\subsection{A priori arguments for ASI being a significant source of existential risk}

Humanity does not have any previous experience dealing with ASI systems. Therefore, we have little evidence we can draw on to estimate how much existential risk would be associated with the creation of an ASI. However, there are still a priori reasons to believe that ASI would be a significant source of existential risk  \cite{withoutspecificcountermeasures}.

What if humanity created an ASI and programmed it with a random goal? What would we expect to happen? Should we expect a positive or negative outcome by default? This section explains why ASI systems are likely to be harmful without careful countermeasures to make them aligned and beneficial.

In a chapter named “Is the default outcome doom?” (p. 140) in \textit{Superintelligence}, Nick Bostrom explains why programming an ASI to be beneficial wouldn’t be easy \cite{superintelligence}. He gives three reasons:

\begin{enumerate}
    \item The orthogonality thesis says that there is no correlation between intelligence and having beneficial goals. Therefore, we cannot expect an AI to acquire beneficial goals simply by increasing its intelligence.
    
    \item It is much easier to program an ASI to have a meaningless goal such as “count digits of Pi” rather than a complex and valuable goal like “achieve human flourishing”.
    
    \item The instrumental convergence thesis says that AIs will have certain sub-goals such as resource acquisition for a wide variety of final goals. For example, if we gave an AI a random goal, it is likely that it would acquire resources to help it achieve its goal. An ASI programmed to maximize the probability of some goal being achieved would be incentivized to pursue extreme and extensive resource acquisition efforts such as building huge numbers of mines and power plants that would make Earth uninhabitable for life. Bostrom calls this particular risk infrastructure profusion.
\end{enumerate}

There are two more reasons why ASI would probably not be aligned by default:

\begin{enumerate}
    \item[4.] To build a beneficial and aligned ASI, one would need to solve the problem of creating ASI in the first place (the AI problem) and also the alignment problem. However, only the AI problem would need to be solved to create a harmful unaligned ASI. If there were a race between organizations creating ASI, the competitor that created the first ASI might then be one that creates an unaligned ASI since doing so would require less effort and time than creating an aligned ASI. Previous technologies such as nuclear power confirm the hypothesis that building safe technology takes longer than harmful technology: the hydrogen bomb was invented in the 1950s but we have yet to discover how to use controlled nuclear fusion reactions to generate electricity.
    
    \item[5.] The set of desirable and beneficial goals we could program an ASI to pursue is a tiny subset of all possible goals. Therefore an ASI with a random goal is very unlikely to be beneficial.
\end{enumerate}

\subsection{Summary of results}

\subsubsection{When will ASI be created?}

\vspace{2mm}

\begin{tabular}{|c|c|c|}
    \hline
    & HLMI / AGI & ASI / TAI  \\
    \hline
    Expert survey 1 & 50\% by 2040 & 10\% by 2042, 75\% by 2070 \\
    \hline
    Expert survey 2 & 50\% by 2066 & \\
    \hline
    Metaculus & 50\% by 2041 & \\
    \hline
    Ajeya Contra & & 50\% by 2040 \\
    \hline
\end{tabular}

\subsubsection{How likely is ASI to cause an existential catastrophe?}

\vspace{2mm}

\begin{tabular}{|c|c|}
    \hline
    Expert survey 1 & 18\% \\
    \hline
    Expert survey 2 & 5\% \\
    \hline
    The Precipice & 10\% in the 21st century \\
    \hline
\end{tabular}

\vspace{5mm}

To answer the question, Ord’s estimate of 10\% this century seems reasonable if we assume that ASI has a high probability of being created in this century and is a significant source of existential risk. The other advantage of this estimate is that since Ord estimated the probabilities of many existential risks, we can compare the size of each risk.

Note that this prediction was made after accounting for the fact that humanity is likely to invest significant effort into reducing existential risk from AI in the 21st century (Ord, 2020, p.170). Therefore, the probability is 10\% conditional on this fact. In a business-as-usual scenario without any effort to align AI, the level of existential risk would probably be far higher than 10\%.

\section{AI timelines and existential risk}

Now that we have estimates for AI risk and other existential risks \cite{theprecipice}, we can analyze how AI timelines would affect AI risk, other existential risks and total existential risk.

In this section, it is important to first introduce the distinction between state risks and step risks (Bostrom, 2014, p.287). State risks are existential risks associated with being in a vulnerable state such as the risk of asteroid strikes. The total amount of state risk accumulates over time. For example, as time goes on the probability of a nuclear war or that earth will be struck by a large asteroid gradually accumulates. Step risks, also known as transition risks, are spikes of existential risk that occur during or immediately after some risky event such as the creation of the first ASI. Unlike state risks, step risks don’t accumulate over time because there is a single risk event. ASI is a step risk because the cumulative level of existential risk would rapidly spike after it is created. Total existential risk would probably then stop increasing because an existential catastrophe would have occurred or because the ASI could prevent all further existential risks. The reason why follows from its definition: an aligned ASI would itself not be a source of existential risk and since it's superintelligent, it would be powerful enough to eliminate all further risks.

\begin{figure}[h]
    \includegraphics[width=12cm]{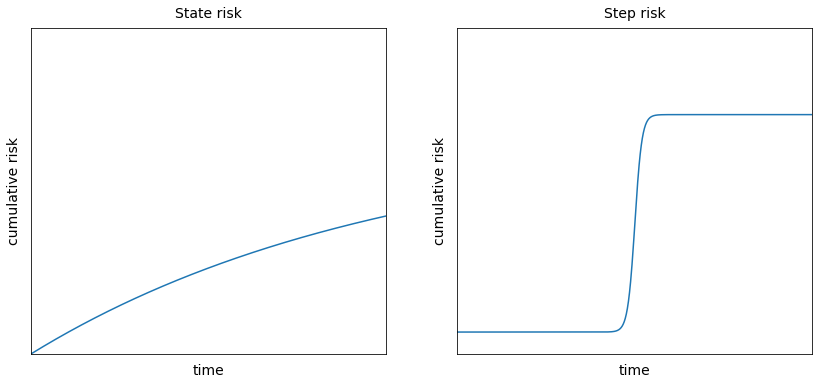}
    \centering
    \caption{State risk vs step risk}
\end{figure}

Delaying the creation of ASI could reduce the amount of step risk resulting from its creation by giving humanity more time to work on the AI alignment problem. But this delay would also increase the accumulation of natural and anthropogenic state risk and make humanity vulnerable to other technological step risks. To study how AI risk and other existential risks interact, it will be useful to first estimate how working on AI alignment would reduce AI existential risk.

\subsection{How quickly would AI alignment work reduce existential risk from ASI?}

\subsubsection{Initial level of AI existential risk}

To estimate how work on the alignment problem would decrease AI existential risk over time, we need to first estimate the initial level of AI existential risk and how it would change as AI alignment expertise accumulates over time from today. A question we can use to estimate the initial level of existential risk is to attempt to answer the question: “If an ASI were created today, what is the probability that it would cause an existential catastrophe?”.

In the previous section, I mentioned Ord’s estimate that AI might have a 10\% probability of causing an existential catastrophe in the 21st century given that humanity has invested significant effort into reducing AI existential risk. But if ASI were created today in 2022, the amount of existential risk would likely be far higher since only a limited amount of AI alignment work has been done so far.

Section 2.5 lists several reasons why an ASI would probably not be beneficial by default. Therefore, if no AI alignment work had been done before the creation of ASI, then the probability of an existential catastrophe happening shortly afterward could be very high - perhaps as high as 90\% or higher.

Given that a limited amount of work has been done on AI alignment as of 2022, this suggests that the level of risk is less than 90\% but more than 10\%. My estimate is 50\% which is very uncertain but as we’ll see, since different existential risks can vary by several orders of magnitude, an estimate within the correct order of magnitude is sufficient for comparing risks \cite{theprecipice}.

\subsubsection{Change in existential risk over time}

The total amount of step existential risk from AI will probably decrease over time as more AI alignment expertise accumulates. But how it would decrease and how fast is unclear as the rate of progress is affected by many factors including:

\begin{description}
    \item[Diminishing returns.] There could be diminishing returns to alignment work as low-hanging fruit are picked. If this factor dominates, we should expect AI risk to decrease rapidly at first and then slow down significantly as the low-hanging fruit are exhausted.
    \item[Change in the number of AI alignment researchers.] Another factor that would affect the rate of progress is the number of AI alignment researchers which has increased significantly over the past several years and is likely to increase in the future as an increasing number of people recognize the importance of AI alignment and as finding a solution seems increasingly urgent. The more AI alignment researchers there are, the faster we can expect AI existential risk to decrease because many researchers can work on subproblems in parallel. Though an increasing number of researchers might also introduce problems such as diseconomies of scale and decreasing productivity per researcher \cite{environmentalresearch}. Even if research progress depends on the quality rather than the number of researchers, a larger pool of researchers would still be beneficial because the maximum level of talent will tend to increase with the number of researchers.
    \item[Change in the difficulty of AI alignment research.] The value and relevance of AI alignment work is likely to increase as the time until the creation of the first ASI decreases. One reason why is that as time goes on AI alignment researchers will be increasingly confident that the AI techniques used in the recent past and in their work will be the same techniques that will be used in the first ASI. Although some AI alignment research is architecture-independent, many AI alignment experiments will require AI researchers to experiment on AI models. If an alignment researcher chooses the wrong architectures for their experiments, their work might be less relevant to the alignment of ASI. Also, as AI capabilities improve over time, we might also expect AI models to understand instructions better and to have greater robustness, generality and other qualities beneficial to AI safety [19].
\end{description}

Each factor affects the shape of the progress curve over time but it’s not clear how the factors would combine. My guess is that the factors combine to create a roughly linear model. Even if progress is unpredictable and not linear, the average rate of progress will still be linear.

The other two variables that need to be estimated are the initial level of AI risk and how fast AI risk would decrease over time. As mentioned in the previous section, my estimate for the initial level of risk was 50\%. The rate of decrease depends on factors described above such as the number of researchers working on the problem and how difficult the alignment problem is to solve. The harder the alignment problem is, the longer it would take to solve.

Two methods of estimating the rate of risk decrease are a bottom-up and a top-down approach. The bottom-up approach involves estimating the rate of progress over a short period of time and extrapolating the rate to a longer time period. The top-down approach involves estimating the total amount of time the alignment problem would take to solve. One could consider the alignment problem to be solved when the total amount of step risk from the transition to the post-ASI era reaches some low probability such as 1 in 1000 which is a risk near zero.

Using the bottom-up method, a reasonable estimate is that each year of alignment research would decrease AI existential risk by approximately 1\% on average which is equivalent to a top-down estimate that the alignment problem would take about 50 years to solve assuming linear progress from an initial risk of 50\%. These estimates are very uncertain but as different existential risks vary by several orders of magnitude, having an estimate within one order of magnitude of the true answer is acceptable.

\subsection{The relationship between AI risk and other existential risks}

Now that we have some idea of how fast progress on the AI alignment problem might occur and or how fast AI existential risk would decrease, we can analyze how delaying or accelerating the advent of ASI would affect the total amount of future existential risk.

Delaying the creation of ASI by a year would give AI alignment researchers an extra year to work on the AI alignment problem which might reduce AI risk by about 1\%. At this point, the obvious strategy for minimizing existential risk is simply to delay the creation of ASI for as long as possible so that we have as much time to solve the alignment problem as possible. The problem is that such a delay also increases other existential risks since humanity would be less effective at mitigating these other risks without the assistance of an aligned ASI.

However, if the total existential risk increase from delaying the creation of ASI is less than the decrease that would result from an additional year of alignment research, then it would be wise for humanity to delay the creation of ASI to minimize existential risk. Conversely, ASI development should be accelerated if each additional year without ASI were associated with an increase in existential risk greater than approximately 1\%.

\subsubsection{Natural state risks}

Let’s first consider the relationship between natural state risk and AI risk. An aligned ASI could reduce or eliminate natural state risks such as the risk from asteroid strikes, supervolcanoes or stellar explosions by devising protective technologies or by colonizing space so that civilization would continue if Earth were destroyed \cite{superintelligence}. But the risk of an existential catastrophe from an asteroid strike is only about 1 in 1,000,000 per century and the total natural risk is about 1 in 10,000 per century (Ord, 2020, p.167). Therefore, if the only source of existential risk other than AI were natural risks, a one-year delay of ASI would decrease AI risk by about 1\% and increase natural existential risk by only 0.0001\%. Therefore, slowing down progress on ASI would, in this scenario, decrease total existential risk and in other scenarios where the total existential risk from factors other than AI is low.

\subsubsection{Anthropogenic state risks}

But there are also anthropogenic state risks such as the risk of nuclear war and climate change. With each passing year, the probability of an all-out nuclear war, catastrophic climate change or resource exhaustion increases. But according to \textit{The Precipice}, these state risks aren’t very high either. The per-century probability of an existential catastrophe from nuclear war is about 1 in 1,000. It’s also 1 in 1,000 for climate change and other environmental damage. And the risk from pandemics is only about 1 in 10,000 \cite{theprecipice}. Again, it wouldn’t be wise to accelerate the development of AI to reduce anthropogenic state risks because the increase in existential risk from accelerating the advent of ASI would be much greater than the decrease in anthropogenic state risk.

The information in the previous two paragraphs is summarized in the following graph and shows that delaying the creation of ASI by one year would cause a net decrease in existential risk if the only other existential risks were natural and anthropogenic state risks including natural risk (from all sources), nuclear warfare, climate change, other environmental damage and natural pandemics.

\begin{figure}[t]
    \includegraphics[width=12cm]{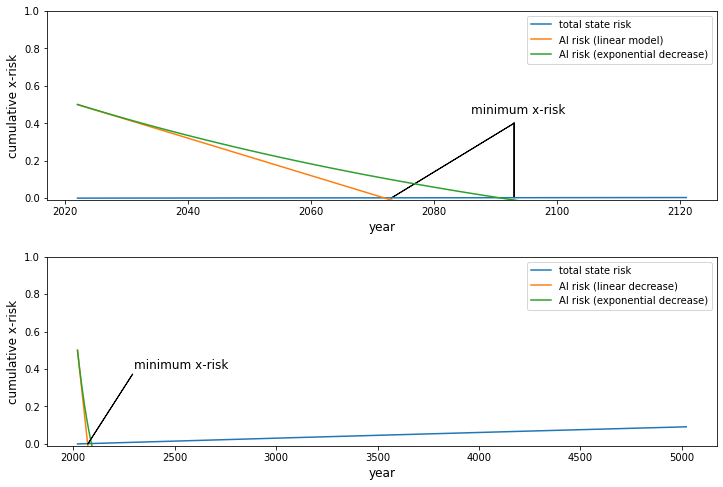}
    \centering
    \caption{AI risk vs total state risk}
\end{figure}

\vspace{5mm}

\subsubsection{Anthropogenic step risks}

The greatest source of existential risk other than ASI is other anthropogenic step risks from risky future technologies such as synthetic biology, nanotechnology and autonomous weapons \cite{theprecipice, globalcatastrophicriskssurvey}. Therefore, the strongest argument for accelerating the advent of ASI is to counter other step risks. Previous dangerous technologies such as nuclear weapons have only been accessible to a few actors. But in the future, the development of technologies such as synthetic biology could make it possible for anyone to cause an existential catastrophe. In such a world, universal surveillance or much better global coordination might be needed to prevent disaster \cite{vulnerable}. Humanity on its own might not be competent enough to safely develop these advanced technologies. The intelligence and power of an aligned ASI would probably increase the probability of humanity passing these risky technological transitions successfully.

How does the sum of other anthropogenic step risks compare to AI risk? According to \textit{The Precipice} \cite{theprecipice}, the risk of an existential catastrophe from engineered pandemics this century is about 1 in 30, for unforeseen anthropogenic risks it’s 1 in 30 and 1 in 50 for other unseen anthropogenic risks. Assuming these risks are independent, the total risk is about 8\% which is similar to the 10\% estimate for AI risk. Therefore, the total amount of AI risk this century might be approximately the same as the sum of all other anthropogenic step risks.

Should progress on ASI accelerate or decelerate given the presence of other step risks? One argument for acceleration is that if ASI were created sooner, we would only need to face AI risk whereas we might need to face the risk of several step risks followed by AI risk if ASI were created later. Therefore it might be desirable to accelerate AI development so that ASI is created before other dangerous technologies \cite{superintelligence}.

\begin{figure}[h]
    \includegraphics[width=12cm]{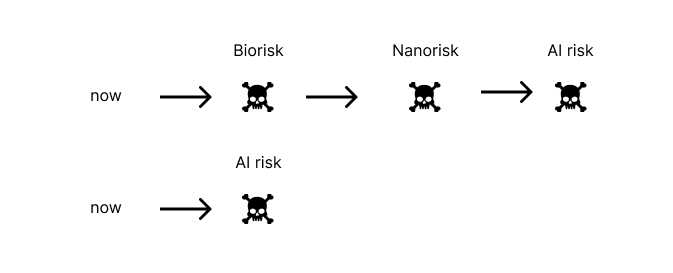}
    \centering
    \caption{reordering strategy}
\end{figure}

The main problem with this ‘reordering’ strategy is that it would reduce the amount of time we have to solve the AI alignment problem which could cancel out its benefits or even increase the total amount of existential risk.

But it’s possible to change the order of arrival without decreasing the amount of time until ASI is created by slowing down the development of dangerous technologies and accelerating the development of beneficial technologies. A similar idea is called \textit{differential technological development} (Bostrom, 2014, p.281):

\begin{quote}
Retard the development of dangerous and harmful technologies, especially ones
that raise the level of existential risk; and accelerate the development of
beneficial technologies, especially those that reduce the existential risks posed by nature or by other technologies.
\end{quote}

\begin{figure}[h]
    \includegraphics[width=12cm]{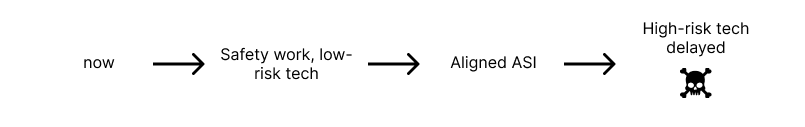}
    \centering
    \caption{differential technological development}
\end{figure}

Differential technological development advocates delaying the development of risky technologies and doing work such as AI safety to reduce existential risk. This strategy would make it possible to delay ASI development to increase the amount of time we have to work on AI alignment without increasing risk from other dangerous technologies. Once an aligned ASI exists, our civilization might then be competent enough to create high-risk technologies such as advanced nanotechnology.

\subsection{Other factors}

The previous sections compared existential risk from AI to other sources of existential risk such as natural and anthropogenic state risks and step risks arising from other risky technologies. But there are factors other than other existential risks that affect AI risk.

\subsubsection{War}

Although a war or nuclear war may be unlikely to cause an existential catastrophe, a war between great powers could still be likely in this century and could increase other existential risks such as AI risk \cite{theprecipice}. Strategies such as differential technological development and international cooperation between countries and AI labs might require a peaceful and favorable geopolitical climate. If war broke out between major powers or even if relations worsened, coordination might become much more difficult. Increasing climate change over the next several decades might also increase conflict and tensions over natural resources such as water or encourage the reckless development of new technologies. A breakdown in international relations might increase technological competition between nations which could increase AI risk. To reduce AI risk, it would probably be desirable to maintain friendly geopolitical relations or implement strategies that require high levels of cooperation while peace lasts.

\subsubsection{Cognitive enhancement}

Over the next several decades it might be possible to create cognitively enhanced humans using genetic engineering that are far more intelligent than unenhanced humans. Although enhanced humans could accelerate progress on the AI alignment problem, they would likely also accelerate progress on the AI problem. Would cognitive enhancement be net positive or negative? Cognitive enhancement might be net positive because whereas we could create ASI using trial and error methods, solving the AI alignment problem might require high levels of foresight, planning and mathematical ability. More intelligent people might also be more likely to have the foresight and abstract thinking necessary to appreciate the importance of AI alignment (Bostrom, 2014, p.289). If cognitive enhancement is beneficial for existential risk reduction, it provides another reason for delaying the creation of ASI so that its creation can be managed by more competent descendants.

\subsubsection{Hardware overhang}

A hardware overhang is a situation where there is an abundance of cheap and powerful computer hardware for running AI. At present, it is probably not economically feasible to train or run a single ASI using current hardware. But as hardware progress continues, it is likely to become possible for one or more companies or nations with large budgets to create an ASI. After that moment, the number of labs that could afford to create an ASI would increase as it became cheaper to do so.

There are several reasons why I think a hardware overhang would be undesirable on balance. A hardware overhang would increase the number of organizations that can afford to create an ASI and might make it possible for small or low-budget teams to create an ASI. Race dynamics would likely be worse with a greater number of teams \cite{superintelligence}. Low-budget labs might not have the resources necessary for safety teams. Coordination would probably be more difficult with a greater number of actors. Large, high-budget labs tend to be more visible which might be useful for inspectors, regulators or researchers analyzing the industry.

Perhaps the worst-case scenario is one where computer hardware is so cheap and powerful that anyone can create an ASI using a single personal computer. In this situation, destruction would be democratized and resulting in a ‘vulnerable world’ \cite{vulnerable} with potentially very high levels of state risk.

If a hardware overhang is undesirable, it may be better if large, well-funded AI teams invested large amounts of money into hardware to be on the frontier of hardware advancement so that if hardware advances to the point where an ASI can be created, the first entity to create an ASI is likely to be a large, well-funded team. Since the first ASI could become a singleton \cite{superintelligence} or perform a pivotal act \cite{pivotal}, once the first aligned ASI is created, the risk of a subsequent unaligned ASI harming the world is probably low.

Although this strategy could reduce AI risk, it might accelerate AI development and reduce the amount of time we have to solve the AI alignment problem. Therefore, as with the reordering strategy, the net effect on the total level of existential risk could easily be negative.

It might be desirable to instead slow down the rate of hardware progress to push back the date when it first becomes possible for a high-budget lab to create an ASI. This strategy would ensure that the entities that are likely to create the first ASI are few, visible and have the resources needed for AI safety without reducing the amount of time humanity has to solve the AI alignment problem.

\section{Conclusions, discussion and recommended action}

Now that we have analyzed how AI risk and other existential risks interact, this section describes some conclusions from the analysis and what we could do to minimize total existential risk.

\subsection{Conclusions}

\subsubsection{How AI timelines affect existential risk}

Delaying the creation of ASI by a year would increase the amount of time humanity has to work on the AI alignment problem by a year. Each additional year of alignment research might decrease AI existential risk by about 1\% which is much greater than the total annual increase in state risk of 0.003\% (300x difference). If state risk were the only source of existential risk other than AI, it would be wise to delay the creation of ASI. Total state risk per century seems to be so low that we could delay the creation of ASI by 1000 years without a significant accumulation of state risk.

However, the existential risk contribution of anthropogenic step risks is much higher. The total amount of step risk from future technologies other than AI could be as high as AI risk \cite{theprecipice, globalcatastrophicriskssurvey} although this quantity is uncertain and it's unclear how it would change over the course of this century. If it’s higher than AI risk or increasing, accelerating the advent of ASI might reduce total existential risk. But if it's relatively low or decreasing, delaying the creation of ASI would reduce existential risk.

\subsubsection{Differential technological development}

A policy of differential technological development might involve delaying the development of all risky technologies such as advanced AI, advanced nanotechnology and synthetic biology while investing in research that reduces existential risk such as AI safety and biosafety. If progress on all risky technologies were slowed down, it would be possible to delay the creation of ASI without simultaneously making humanity vulnerable to other step risks such as the invention of advanced nanotechnology.

It may be difficult to achieve the level of coordination necessary to implement differential technological development. However, even if these other dangerous technologies could not be delayed, since AI risk is probably the highest step risk, an additional year of AI delay might still be beneficial for reducing total existential risk. But unlike the comparison of AI risk and total state risk, it probably wouldn’t be wise to delay ASI development for hundreds or thousands of years because the total amount of step risk from technologies other than AI is generally far higher than the total amount of state risk. Assuming there are diminishing returns to AI alignment research, at some point we would expect the marginal decrease in existential risk from AI alignment research per year to be exceeded by the annual existential risk increase from the possibility of some other dangerous technology being created such as self-replicating nanotechnology.

\subsubsection{Other factors}

\begin{description}
    \item[War, climate change and international cooperation.] Although the direct existential risk contribution from state risks such as war and climate change are relatively low, these risks could weaken civilization or worsen the political climate and make it more difficult to implement cooperative actions such as an agreement to avoid arms races or weaponization. A worse political climate or poorer world might also reduce our ability to implement differential technological development and indirectly cause a significant increase in existential risk. For example, it might be more difficult to advocate slowing down progress on potentially beneficial but risky technologies such as nanotechnology or AI if humanity is suffering from the effects of severe climate change and desperate to try anything that might solve the problem.
    \item[Hardware overhang.] We have also seen that it might be better if the development of ASI is carried out by as few actors as possible to reduce the severity of race dynamics and avoid creating a vulnerable world. Since a hardware overhang would increase the number of actors that can create an ASI, it would probably be undesirable. It might be desirable for large AI labs to invest heavily in hardware so that they are capable of creating an ASI before large numbers of other less well-funded teams or individuals.  While this strategy might accelerate AI research, it could also be implemented while slowing down AI research if progress on computer hardware were slowed down.
    \item[Cognitive enhancement.] Although cognitive enhancement would probably accelerate AI progress, progress on AI alignment might benefit more from cognitive enhancement than AI progress. Also, a cognitively enhanced population might be more likely to appreciate the importance of the AI alignment problem.
\end{description}

\subsection{Recommended actions}

We have seen how AI timelines affect existential risk and how various existential risks and other factors can interact in complex ways. Consequently, there is substantial uncertainty about which actions are net beneficial or harmful. Missing one or more crucial considerations could lead to unintended harm. An action that seems beneficial could actually be harmful or vice-versa.

The solution to the problem is to take robustly positive action \cite{superintelligence} which is action that is very likely to be beneficial across a wide range of possible scenarios. Some strategies such as accelerating AI research so that ASI is created before other dangerous technologies are not robustly positive because there is a significant chance that such a strategy could have a net negative effect. Conversely, slowing down progress on AI could cause net harm if it causes a hardware overhang or if the delay leads to the accumulation of risk from other dangerous technologies.

\subsubsection{Robustly positive actions}

\begin{description}
    \item[Existential risk and AI strategy research.] Although previous work and this report arrive at some useful conclusions, there is still substantial uncertainty about existential risk and which actions would be useful to reduce it. Since some philosophical theories such as longtermism say that existential risk reduction has extremely high value, further research would be valuable. Also, until relatively recently research on existential risk and AI strategy has been neglected. Therefore, it is likely that there are still important insights that could be discovered in the future. One goal of strategic analysis is to find crucial considerations (Bostrom, 2014, p.317) which are:
        \begin{quote}
        ideas or arguments with the potential to change our views not merely about the fine-
        structure of implementation but about the general topology of desirability. Even
        a single missed crucial consideration could vitiate our most valiant efforts or render them as actively harmful as those of a soldier who is fighting on the wrong side.\cite{superintelligence}
        \end{quote}
    \item[AI safety research.] Although creating advanced AI would benefit the world in many ways, the potential risks of advanced or superintelligent AI seem high. Therefore, advancing AI capabilities does not qualify as a robustly positive action. In contrast, AI safety or alignment research is likely to have robustly positive value because AI safety research decreases existential risk and is unlikely to backfire and increase existential risk. It would also be valuable to increase the number of AI safety researchers to increase the probability of the AI alignment problem being solved before the first ASI is created.
    \item[Research on other existential risks.] There are many sources of existential risk other than AI. Therefore humanity should work on reducing all existential risks to minimize total existential risk. Reducing other existential risks might also allow humanity to safely delay the creation of ASI if more time were needed to solve the AI alignment problem.
    \item[Creating a favorable global environment.]  In addition to other existential risks, other factors such as war, climate change and poverty might increase existential risk indirectly by reducing global cooperation and coordination, increasing competition and increasing the risk of hasty and reckless action. Therefore, actions such as maintaining international peace and avoiding climate change probably have robustly positive value \cite{theprecipice}.
    \item[Raising awareness of existential risk.]  The idea of existential risk and its importance has only been recently recognized \cite{existentialriskhistory}. Increasing awareness of existential risk and related ideas especially among key decision makers would increase the chance that humanity takes wise actions that reduce existential risk in the future.
\end{description}

\subsubsection{Other actions}

I’ve listed other actions here that might have positive value but that I’m not confident would have robustly positive value.

\begin{description}
    \item[Cognitive enhancement.] As mentioned earlier, cognitive enhancement could reduce AI risk by accelerating progress on the AI alignment problem more than the AI problem if solving the alignment problem is more dependent on difficult foresight and deductive reasoning rather than experience and improvement by trial and error. However, creating cognitively enhanced human researchers could be difficult. Also, if AI progress is fast, the first generation of enhanced researchers is unlikely to be ready before the first ASI is created. Nevertheless, there are alternative actions that would have similar benefits to cognitive enhancement that could be done today:
    \begin{description}
        \item[Making AI safety problems more concrete.] Humanity is less likely to solve the AI alignment problem in time if we can only do so via difficult first-principles thinking or if we can only solve the problem in a single attempt. Creating concrete AI safety problems in the form of challenges or benchmarks would make it possible to make incremental progress or make progress by trial and error. Therefore, the same experimental methods that have fueled progress on AI could be used for AI safety which might help ensure that AI safety keeps up with other AI progress.
        \item[Increasing the number of AI alignment researchers.] The number of exceptionally talented researchers working on AI safety is likely to increase as the number of AI safety researchers increases.
        \item[Breaking hard problems into subproblems.] Progress on hard problems could be accelerated by breaking them into subproblems and assigning one or more researchers to work on each subproblem. Using this strategy it might be possible for a large number of moderately talented researchers to match the performance of a few exceptionally talented researchers.
    \end{description}
    \item[Avoiding a hardware overhang.] As mentioned earlier, a hardware overhang could increase AI risk by increasing the number of actors capable of creating an ASI and possibly worsening race dynamics. If leading AI companies invested heavily in AI hardware, a hardware overhang would be less likely but such a strategy could accelerate the advent of ASI. Slowing down hardware progress might be beneficial but doing so would be difficult since organizations such as chip manufacturers have strong incentives to improve hardware. One proposal that would be more feasible is limiting access to large amounts of AI hardware to only trustworthy organizations. For example, AI hardware providers such as AWS could restrict AI hardware access to untrustworthy entities or entities that don’t demonstrate a strong commitment to AI safety \cite{parler}.
\end{description}

\printbibliography

\end{document}